\documentclass[aps,twocolumn, superscriptaddress,groupedaddress,nofootinbib]{revtex4-1}
\usepackage[dvipsnames]{xcolor}
\usepackage[utf8]{inputenc}
\usepackage{url}
\usepackage{amsmath}
\usepackage{tikz}
\usepackage{amsfonts}
\usepackage{amssymb} 
\usepackage{bm}
\usepackage{graphicx}
\usepackage{array}
\usepackage{lipsum}
\usepackage{float}
\usepackage{multirow}
\usepackage{hhline}
\usepackage{tabularx}
\usepackage{subfig}
\usepackage{graphicx}
\usepackage{afterpage}
\usepackage{epstopdf}
\usepackage{MnSymbol}
\usepackage[utf8]{inputenc}

\usepackage{amsthm}

\makeatletter
\renewcommand\@makecaption[2]{%
  \par
  \vskip\abovecaptionskip
  \begingroup
   \small\rmfamily
    \begingroup
     \samepage
     \flushing
     \let\footnote\@footnotemark@gobble
     \@make@capt@title{#1}{#2}\par
    \endgroup
  \endgroup
  \vskip\belowcaptionskip
}
\makeatother

\definecolor{cornellRed}{HTML}{B31B1B}

\def\cV{\mathcal{V}}

\def\gagg{g_{a\gamma\gamma}}

\begin{document}
\title{On String Theory Expectations for \\ Photon Couplings to 
Axion-Like Particles}
\author{James Halverson, Cody Long, Brent Nelson, and Gustavo Salinas}
\affiliation{Department of Physics, Northeastern University \\ Boston, MA 02115-5000 USA} 

\date{\today}

\begin{abstract}
ALP-photon couplings are modeled in
large ensembles of string vacua
and random matrix theories. In all cases, the effective
coupling increases polynomially
in the number of ALPs, of which
hundreds or thousands are expected
in the string ensembles, many of which
are ultralight. The expected value of the couplings $g_{a\gamma\gamma}\simeq 10^{-12}\text{GeV}^{-1} - 10^{-10}\text{GeV}^{-1}$
provide viable targets for future x-ray
telescopes and axion helioscopes, and
in some cases are already in tension
with existing data.
\end{abstract}

\maketitle

\section{Introduction}

If string theory is the correct theory of quantum gravity, one of its vacua must realize the photon of classical electromagnetism. Uncharged spin zero particles
must couple to the electromagnetic
field strength, since 
all couplings in string theory
are determined by vacuum expectation
values (VEVs) of scalar fields. 
This applies not only to the usual
parity even operator $F_{\mu\nu}F^{\mu\nu}$, but also the parity odd operator,
requiring the existence of a coupling
\begin{equation}
\mathcal{L} \supset -\frac{1}{4}g_{a\gamma\gamma}\, a F_{\mu\nu}\tilde F^{\mu\nu}
\end{equation}
in the effective Lagrangian, where $\tilde F^{\mu\nu} = \epsilon^{\mu\nu\rho\sigma}F_{\rho\sigma}$.,
The pseudoscalar $a$ is an axion-like
particle (ALP), which is not necessarily
the QCD axion, and it may be a non-trivial linear combination of the hundreds or thousands of ALPs expected
from studies of string vacua.

Numerous ground-based experiments \cite{Collaboration2017,Salemi:2019xgl,Armengaud:2019uso,Bogorad:2019pbu} and satellite observations \cite{Wouters:2013hua,Berg:2016ese,Marsh:2017yvc,Conlon:2017qcw,Day:2019ucy,Reynolds:2019uqt,Conlon:2017ofb}
place constraints on ALP-photon interactions, probing widely different regimes for the axion mass
$m_a$ and coupling strength $g_{a\gamma\gamma}$.
Existing
limits are already remarkable, within a few orders of magnitude of  GUT scale decay constants $f_{a\gamma\gamma} \equiv g^{-1}_{a\gamma\gamma}$.
However, the exclusions depend critically on whether the ALP is assumed to be a sizable fraction of the
dark matter and also experimental limitations affecting the accessible mass range. Since dark matter in
string theory is often multi-component (e.g., \cite{Dienes:2011ja,Halverson:2016nfq}) and
many ALPs are expected to be ultralight, but not yet in any fixed mass window, we will focus on analyses relevant for experiments that do not require ALP dark matter and only set limits below a fixed mass threshold.

For such experiments, the strongest bounds on
the coupling arise at low mass, $m_a\leq 10^{-2}\, \text{eV}$, where results \cite{Collaboration2017} from the axion helioscope CAST require $g_{a\gamma\gamma}\leq 7\times 10^{-11} \, \text{GeV}^{-1}$. The CAST result already significantly outperforms projected collider bounds for $m_a\simeq$ GeV scale ALPs, leading us to focus on the
low mass range. For even lower masses, $m_a\leq 10^{-12}\, \text{eV}$, observations from the x-ray telescope Chandra 
require  $g_{a\gamma\gamma}\leq 8\times 10^{-13} \, \text{GeV}^{-1}$ \cite{Reynolds:2019uqt}. The future helioscope
IAXO and satellite STROBE-X are projected to probe
$g_{a\gamma\gamma}\simeq 2\times 10^{-12}\,\text{GeV}^{-1}$ for $m_a\leq 10^{-2}\, \text{eV}$ \cite{Armengaud:2019uso} and $g_{a\gamma\gamma}\simeq 8\times 10^{-14}\,\text{GeV}^{-1}$ for $m_a\leq 10^{-12}\, \text{eV}$ \cite{Ray:2019pxr}, respectively.

The primary focus of this work is to understand
the dependence of $g_{a\gamma\gamma}$ on the number $N$ of ALPs in both models of string vacua and random matrix effective field theories. Though in the $N=1$ case (using mild assumptions described in the text) we have
\begin{equation}
\text{Single ALP:} \qquad \qquad \gagg = \frac{1}{\sqrt{3} M_p},
\end{equation}
independent of the details of the compactification geometry (including the string scale), it is reasonable to imagine that the introduction of additional ALPs will increase the effective
coupling to photons.  Concretely, we study whether the bulk of the $\gagg$ distribution at large $N$ is in a range
probed by current or future experiments. This is of interest because most vacua are expected to arise at the
largest values of $N$ afforded by the given ensemble. We will refer to the mean of the distribution at this value of $N$ as the expected value
of $\gagg$.

Our main result is that in all string ensembles and random matrix theories that we study, $\gagg$ increases polynomially in $N$. We perform detailed studies of two string geometry ensembles that we refer to as the tree ensemble and the hypersurface ensemble, using the
mass threshold $m_a\leq 10^{-12}\, \text{eV}$ so that our results can be compared to numerous experiments. In the tree ensemble
at the expected value of $N=2483$, we find that the mean of the projected $\gagg$ distribution is $\gagg = 3.2\times 10^{-12} \, \text{GeV}^{-1}$. In the hypersurface
ensemble $N=491$ is expected, and we find the expected value $\gagg = 2.0\times 10^{-10}\,\text{GeV}^{-1}$. We also study the F-theory geometry with the
most flux vacua \cite{Taylor:2015xtz}, which is very constrained due to the existence of only a single divisor that can support the Standard Model, given our assumptions; it yields $\gagg = 3.47 \times 10^{-12}\,\mathrm{GeV}^{-1}$. All of these couplings are in range
of future experiments, and in some cases are already in tension with data; see the discussion. 
Furthermore, by removing the mass threshold the expected value of $\gagg$ does not change significantly, implying that ALPs in string ensembles that we study will not be seen in
searches at LHC or future colliders.
The random matrix results suggest that $\gagg$ should increase significantly with $N$, 
even if our vacuum does not arise in one of the studied ensembles.

\vspace{.2cm}

We emphasize at the outset that we are \emph{modeling} ALP-photon couplings using data from string theory. A complete calculation, with engineered Standard Models and
full moduli stabilization, is computationally intractable given current techniques, which can 
be formalized in the language of computational complexity \cite{Halverson:2018cio}. At small $N$, however, more complete calculations can be performed, including partial moduli 
stabilization; see, e.g., \cite{Cicoli:2012sz}. Instead, we model ALP-photon couplings by intelligently sampling the Calabi-Yau moduli space and utilizing knowledge of how realistic
gauge sectors arise, without trying to concretely engineer them or stabilize moduli. This allows for the study of large ensembles at large $N$,
but also motivates further studies once techniques for the complete calculations become available. The models that we use are described thoroughly in the text and accurately summarized in the discussion.

This paper is organized as follows. In section \ref{sec:eft} we discuss ALP-photon couplings
from the perspective of string theory. In section \ref{sec:ensembles} we introduce the ensembles of
string compactifications and random matrix effective theories.  In section \ref{sec:ens} we present the $g_{a\gamma\gamma}$ distributions computed in the ensembles. We review constraints and projections of existing and proposed experiments in section \ref{sec:experiments} and 
discuss our results in light of them in section \ref{sec:discuss}. Interesting future directions are also discussed.

\section{ALP-Photon Couplings in String Theory \label{sec:eft}}

Low-energy effective field theories from string theory regularly have a large number of ALPs, and under certain assumptions related to control of the theory
many of them are very light, as we will soon discuss. The Lagrangians of interest, focusing on the ALP sector, take the form
\begin{align}\label{eqn:cannorm1}
& \mathcal{L} = - \frac{1}{4} F^{\mu \nu}F_{\mu \nu} -\frac{1}{2}\delta_{ij} (\partial^\mu \phi^i)(\partial_\mu \phi^j) \nonumber \\
&  - m_i^2 (\phi^i)^2 - \frac{1}{4}c_i \phi^i \tilde{F}^{\mu \nu}F_{\mu \nu} \, .
\end{align}
Here $F_{\mu \nu}$ is the electromagnetic field strength, with $\tilde{F}^{\mu \nu} = \epsilon^{\mu\nu\gamma\sigma} F_{\gamma \sigma}$, and $\phi^i$ are the ALPs, with masses $m_i$.
From a low-energy perceptive these parameters are generally unconstrained; however, we will find that UV structure from string theory actually constrains the low-energy physics, and introduces correlations and patterns, that will result in interesting structure in both the ALP masses and ALP-photon couplings.

For concreteness, we will study the dependence of ALP-photon couplings on the number of ALPs $N$ in string compactifications and random matrix models.  In the latter context,
we consider compactifications of type IIB string theory/F-theory on a suitable K\"ahler manifold $B$, which yields a 4d $\mathcal{N} = 1$ EFT.  To date, such compactifications encompass the largest known portion of the $\mathcal{N} = 1$ landscape~\cite{Taylor:2015ppa, Taylor:2015xtz, Halverson:2017ffz, Halverson:2017vde, Taylor:2017yqr}. Some of the most generic phenomena in this region, such as large gauge sectors and numbers of ALPs,
correlate strongly with moving away from weakly coupled limits \cite{Halverson:2016vwx}. The ability to make such statements away from weak string coupling
relies critically on the holomorphy implicit in algebraic geometry. 

The data of such an F-theory compactification, in addition to the choice of $B$, is an elliptic fibration over $B$ that encodes the data of the gauge group, see~\cite{Weigand:2018rez} for further details.
The ALPs $\theta^i$ that we study arise from the dimensional reduction of the Ramond-Ramond four-form $C_4$, and become the imaginary parts of complexified K\"ahler moduli, written as
\begin{equation}
T^i = \int\limits_{D_i} \left(\frac{1}{2} J \wedge J + i\, C_4 \right) \equiv \tau^i + i\, \theta^i \, .
\end{equation}
Here the $D_i$ are a basis of divisors (4-cycles) in $B$, numbering $h^{1,1}(B)$, and $J$ is the K\"ahler form on $B$, which can be written in terms of parameters $t_i$ as $J = t_i \omega^i$, with $\omega^i \in H^{1,1}(B)$. The $\tau^i$ parametrize the volumes of the divisors in $B$. A typical $B$ has $h^{1,1}(B) \sim \mathcal{O}(10^3)$~\cite{Halverson:2017ffz, Halverson:2019kna}; that is, in these
ensembles, thousands of ALPs are expected.

In addition to the many ALPs, EFTs arising from string theory often contain many gauge sectors. Throughout
we will take $N$ to be the number of ALPs.  In the context of F-theory, our focus will be on when this gauge sector, indexed by $\alpha$, is supported on a stack of 7-branes wrapping a cycle $Q^{\alpha}$, or from a non-trivial Mordell-Weil group element of the elliptic fibration $Q^{\alpha}$.
The ALP-gauge portion of the EFT takes the form
\begin{align}\label{eqn:eff}
& \mathcal{L}  = -M_p^2 K_{ij} (\partial^\mu \theta^i)(\partial_\mu \theta^j) - V(\theta)  \nonumber \\ 
& - \sum\limits_\alpha Q^{\alpha}_{\, i} \left( \tau^i G_\alpha^{\mu\nu}G_{\alpha \mu \nu} + 
 \theta^i \tilde{G}_\alpha^{\mu\nu}G_{\alpha \mu \nu}\right) \, .
\end{align}
Here $M_p$ is the Planck mass, $V(\theta)$ is the non-perturbative ALP potential, and $K_{ij}$ is the metric on moduli space, which at tree level is derived from the K\"ahler potential $\mathcal{K} = -2\, \mathrm{log} \cV$, where $\cV$ is the volume of $B$, which is expressed as
\begin{equation}
\cV = \int\limits_B J \wedge J \wedge J = \frac{1}{6}\kappa^{ijk}t_i t_j t_k\, ,
\end{equation}
and $\kappa^{ijk}$ are the triple intersection numbers of $B$.

To justify our use of the tree-level K\"ahler potential, we note that F-theory can be viewed as dimensional reduction of type IIB SUGRA in the presence of a spatially-varying axio-dilaton. The type IIB SUGRA action is unique up to field redefinitions and higher-derivative corrections, and so any physical correction to the 4d EFT must come with additional powers of appropriate volumes of cycles. At large enough volume we therefore expect that using $\mathcal{K} = -2\, \mathrm{log}\cV$ should be a good approximation to the K\"ahler potential. It is not known how large is large enough in this case, but since the volumes in compactifications with many cycles tend to be very large~\cite{Demirtas:2018akl}, we believe this approximation should be valid for our study.

The leading-order K\"ahler potential is independent of the ALPs, so one can move to a canonically normalized frame in which the ALP mass matrix is diagonal. Let $F_{1\mu\nu} \equiv F_{\mu \nu}$ be the electromagnetic field strength, corresponding to a homology class $Q$ (subtleties associated with electroweak symmetry breaking will be discussed momentarily). The Lagrangian of interest then takes the form  
\begin{align}\label{eqn:cannorm}
& \mathcal{L} = -\frac{1}{2}\delta_{ij} (\partial^\mu \phi^i)(\partial_\mu \phi^j)  - \frac{1}{4} F^{\mu\nu}F_{ \mu \nu}   \nonumber \\ 
& -\frac{1}{4} c_{i}\phi^i \tilde{F}^{\mu\nu}F_{ \mu \nu} - m_i^2 (\phi^i)^2 \, .
\end{align}
In terms of geometric data, we can write~\cite{Halverson:2019kna}
\begin{equation}\label{eq:gcoups}
|\vec{c}| = \frac{\sqrt{Q \cdot K^{-1} \cdot Q}}{2 M_p Q \cdot \tau} \, , 
\end{equation}
where $K^{-1}$ is the inverse K\"ahler metric on field space. Eq.~\ref{eq:gcoups} is independent of homogeneous scaling in the K\"ahler cone $J \rightarrow \lambda J$, and therefore only depends on the angle in the K\"ahler cone. 

It is interesting to note that if $N = 1$ then this coupling is fixed to be $c = 1/(\sqrt{3} M_p)$ 
independent of the details of the geometry~\cite{Demirtas:2019lfi}. This is important because in the case of a single ALP it determines the photon
coupling, $\gagg\simeq 1/M_p$, which is well below current and (projected) future experimental bounds. We will investigate whether large $N$ effects lead to a significant enhancement of the coupling. Note that if the masses of all the $\phi^i$ that appeared non-trivially in $c_i \phi^i$ were the same, then $|c|$ would simply be the coupling of the canonically normalized field $\varphi \equiv c_i \phi^i/|c|$ to $F_{\mu\nu}$. In general this will not be the case; however, in string compactifications with $N$ large we expect that some of the ALPs will be essentially massless, which we will now review. 

ALP masses depend critically on the fact that, in the absence of sources,
the $\theta^i$ enjoy a continuous shift symmetry to all orders in perturbation theory, broken to a discrete shift symmetry by non-perturbative effects. 
In particular, the superpotential $W$ is known to receive corrections from stringy instantons and/or strong gauge dynamics that generate masses for the ALPs~\cite{Witten:1996bn}. Due to the shift symmetry of the ALP and the holomorphy of the superpotential, any non-perturbative contribution to the superpotential takes the schematic form
\begin{equation}\label{eqn:super}
\Delta W \sim e^{-2 \pi \tilde{Q}_i (\tau^i + i \theta^i)}\, ,
\end{equation}
for rational $\tilde{Q}_i$. Therefore, for a given $\theta^i$, if the corresponding $\tau^i$ is very large, then any non-perturbative contribution to $W$ involving $\theta^i$ will be negligible, and $\theta^i$ is expected to be essentially massless (the same is not true of the $\tau^i$, as they do not enjoy a similar shift symmetry and can receive masses via perturbative corrections to the K\"ahler potential~\cite{Cicoli:2008va, Cicoli:2016chb}). 
The central observation of~\cite{Demirtas:2018akl} is that in compactifications with large $N$, the region of moduli space where the EFT is expected to be valid (known as the stretched K\"ahler cone) is quite narrow. Restricting to the stretched K\"ahler cone for the sake of control forces some of the $\tau^i$ to be very large, which in turn forces some of the $\theta^i$ to be extremely light. This is the key result that (in this context) puts numerous ALPs in the sub-eV mass range relevant for the experiments we will discuss. In this work we make the technical assumption that the Standard Model sector does not generate a large mass term for the ALP ($\gtrsim 10^{-12} \mathrm{eV}$). This assumption is quite mild given the scales of the potential generated by $SU(2)_L$ instantons, as well as current expectations for the QCD axion mass~\cite{diCortona:2015ldu, McLerran:2012mm, McLerran:2014daa}.

If the ALP masses are much lower than the typical energies of an experiment then they can be safely neglected, and the ALPs taken to be massless. We will assume that the typical experiment energies are $\gtrsim 10^{-12} \mathrm{eV}$, and so any ALP with a mass $\ll 10^{-12}\, \mathrm{eV}$ can be treated as massless. In this case we can
define a single ALP $a \equiv c_i \phi^i/|c|$ that couples to $F_{\mu \nu}$, with strength $\gagg$. Here the sum includes only the $\phi^i$ that have negligible mass. The relevant terms in  Eq.~\ref{eqn:cannorm} become
\begin{align}\label{eqn:imp}
& \mathcal{L} \supset -\frac{1}{2} (\partial^\mu a)(\partial_\mu a) - \frac{1}{4}  F^{\mu\nu}F_{\mu \nu} \nonumber \\
& -\frac{1}{4} \gagg a \tilde{F}^{\mu\nu}F_{\mu \nu}\, ,
\end{align}
where
\begin{equation}\label{eqn:cncoup}
\gagg = \sqrt{\sum\limits_i c_i^2}\,,
\end{equation}
in terms of the couplings $c_i$ in Eq.~\ref{eqn:cannorm}. Note that we can always redefine our ALP $a$ to make $\gagg$ non-negative, as we do henceforth. 

It is useful to express Eq.~\ref{eqn:cncoup} in terms of the original geometric quantities that appear in Eq.~\ref{eqn:eff}. To do so, we determine the massless axions by finding the linear combinations of axions that receive a non-negligible mass term (see below for a discussion), in the canonically-normalized frame. Let the massive axions be specified by a (generally non-full rank) matrix $M^{a}_{i}$, such that $\theta^a \equiv M^a_i \theta^i$ receives a non-negligible superpotential contribution in Eq.~\ref{eqn:super}. We can move to a canonically normalized frame by writing $K = S^T\cdot  f\cdot  f\cdot S$, where $S$ is a matrix of the orthonormal eigenvectors of $K$, and $f$ is a diagonal matrix of the square-roots of the eigenvalues. The massless axions (in the canonically-normalized frame) are then specified by the matrix
\begin{equation}
D = \mathrm{Ker}(M \cdot S^T \cdot f^{-1})\,;
\end{equation}
i.e. the massless linear combination of canonically normalized axions is encoded in the rows or columns.

We can express these axions in the geometric $\theta$-basis by writing
\begin{equation}
Q^M = D \cdot f \cdot S\, ,
\end{equation} 
where the superscript refers to ``massless'', and we may therefore rewrite $\gagg$ in Eq.~\ref{eqn:cncoup} as
\begin{equation}
\label{eqn:gaggeqn}
\gagg = \frac{\sqrt{Q^M \cdot K^{-1} \cdot Q^M}}{2 M_p Q \cdot \tau} \, ,
\end{equation}
for fixed gauge group specified by the homology class $Q$.

Finally, we comment briefly on the string scale $M_s \simeq M_p / \sqrt{\cV}$. For reasons that we will discuss, obtaining control over the string effective theory at large $N$ requires going to regions
in K\" ahler moduli space where a non-trivial number of four-cycle volumes are large. This correlates strongly with a large overall volume $\cV$, which generally gives rise to an intermediate string scale at large $N$ in our ensembles, $M_s \simeq 10^{12} - 10^{15} \, \text{GeV}$. However, since it is
divisor volumes that more readily appear in our $\gagg$ calculations and the geometry does not
necessarily have the Swiss cheese property, we prefer to think of the relationship between
$M_s$ and $\gagg$ as correlative, rather than causal. It would be interesting to explore this
further in future work.

Our goal is to to compute $\gagg$ in large ensembles of string compactifications, as well as random matrix models, in order to model the distribution of such couplings that one expects from string theory, and understand the behavior as the number of ALPs $N$ grows large.

\section{Ensembles \label{sec:ensembles}}

Having reviewed the EFTs expected from string theory,
the specific contexts in which we will study them, and the presence of ultralight ALPs, we now introduce the ensembles in which we perform
this study.

\subsection{The Tree ensemble}\label{sec:tree}
 F-theory is a non-perturbative generalization of type IIB string theory that allows for regions of strong string coupling, and has a more general gauge spectrum than its IIB counterpart. In F-theory, the internal space $B$ for compactification determines a minimal (geometric) gauge structure from non-Higgsable 7-branes, whose presence requires no tuning in complex structure moduli space.  

 In~\cite{Halverson:2017ffz} a lower bound of the number of bases $B$ suitable for F-theory compactifications was determined to be $4/3 \times 2.96 \times 10^{755}$ via the discovery of a construction algorithm for an ensemble of $B$ known as the Tree ensemble.
 This ensemble represents a large graph,
 where nodes are geometries and edges
 are simple topological transitions (known as blowups) that may
 take place between the geometries. 
 More specifically, each node contains 
 a parametric family (the complex
 structure moduli space) of geometries that
 have the same topological type, but for 
 some representatives of the family
 the space becomes singular enough to
 allow for a topological transition. This
 means that at leading order in the physics,
 the space can be explored by movement
 along flat directions of the scalar
 potential to reach the singular point,
 from which a transition may be made to
 a different topological type (node) of
 geometry and accordingly a different
 EFT. Mathematically, this process
 may be done continuously in moduli space,
 as theorems relating canonical singularities
 and the Weil-Petersson metric ensure that
 the paths through the graph are at
 finite distance in moduli space. See \cite{Hayakawa, Wang,Halverson:2017ffz, Carifio:2017nyb} for further discussion. 

 The Tree ensemble, while enormous in size, has certain tractable aspects, due to detailed understanding of the construction algorithm. In particular, the  geometric gauge group can be determined to high accuracy. Such a set provides a rich ensemble in which to address questions about distributions of effective field theories in string theory. The Tree ensemble is constructed by starting with a weak Fano toric variety $B$, and performing blowups of $B$ that satisfy sufficient conditions to remain at finite distance in moduli space. Such blowups can be performed over toric points or curves, and each sequence of blowups of a particular toric point or curve in $X$ is called a tree.
A typical EFT from the Tree ensemble has a minimal gauge group of the form
\begin{equation}\label{eqn:mingauge}
G \geq E_8^{10} \times F_4^{18} \times U^9 \times F_4^{H_2} \times G_2^{H_3} \times A_1^{H_4}\, ,
\end{equation}
where $U$ is a $B$-dependent gauge group, and $H_i$ are computable $B$-dependent integers that are almost always non-zero. Other gauge groups can be tuned, but the group given in Eq.~\ref{eqn:mingauge} is required by the geometry (though some of the gauge factors could be broken by the introduction of $G_4$ flux). 

In addition, an overwhelming fraction of EFTs from the Tree ensemble have a large number of ALPs. The expected number of ALPs is $N = 2483$. This value is determined by a bubble cosmology model on the Tree ensemble, where the vacuum transitions are modeled by the topological transitions between the geometries~\cite{Carifio:2017nyb} (drawing from a flat distribution gives similar results, with the preferred value of $N = 2015$).

From the expected gauge structure of the Tree ensemble discussed in the previous section it is clear that there are in principle many ways to realize the Standard Model in the Tree ensemble. In addition to the minimal gauge structure given in Eq.~\ref{eqn:mingauge} one can tune additional gauge groups, and additional $U(1)$ factors could be realized by sections in the elliptic fibration.

For understanding the couplings of the ultralight ALPs to the photon, the relevant gauge sector is $SU(2)_L \times U(1)_Y$. Since pure abelian factors have not been studied in the Tree ensemble, we will consider two cases of gauge sectors. First, one could try
to realize $SU(2)_L \times U(1)_Y$ by embedding it in a larger non-Abelian gauge group, for instance a GUT, in which case we will model the coupling $\gagg$ directly as the coupling of the ALP to the larger group, assuming it breaks to the Standard Model in one of the canonical ways. The second case is realizing $SU(2)_L$ directly from the geometrically determined $SU(2)$ gauge symmetry on the seven-brane, in which case we will compute the contribution to $g_{a\gamma\gamma}$ from $SU(2)_L$. The cases are summarized as:
\begin{enumerate}\label{en:options}
\item $G =  SU(2)_L$ arises directly on a seven-brane,
\item $G$ is a geometric gauge group on a seven-brane such that $G \rightarrow SU(2)_L \times U(1)_Y \times G^{'}$, but $G^{'} \nsupset SU(3)$,
i.e. QCD comes from a different seven-brane.
\item $G$ is a gauge group on seven-brane such that $G \rightarrow SU(2)_L \times U(1)_Y \times SU(3)$; this includes some common GUT scenarios.
\end{enumerate}
In the last option the ALP linear combination that couples to $SU(2)_L$ is the QCD axion, while in the first two options this is not necessarily the case.
Since all three possibilities lead to similar results, the forthcoming plots of ALP-photon couplings take into account all three.

In case $1$, where we compute the contribution to $g_{a\gamma\gamma}$ from $SU(2)_L$, which we denote $g_{aWW}$, the coupling to the physical photon is computed as
\begin{equation}
\gagg = g_{aWW} \mathrm{sin}^2 \theta_w + g_{aYY}\mathrm{cos}^2\theta_w\, ,
\end{equation}
where $\theta_w$ is the Weinberg angle ($\mathrm{sin}^2 \theta_w \simeq 0.23$), and $g_{aYY}$ is the coupling to $U(1)_Y$.
In only this case it is possible that $g_{aYY}\neq g_{aWW}$. However, as modeling the hypercharge in our ensembles is non-trivial, we 
simply assume that the hypercharge contribution to $\gagg$ does not lead to significant cancellation. This would require $g_{aWW}$ and $g_{aYY}$ to be of 
similar order of magnitude and opposite sign. As long as this is not the case, which we find plausible, $|g_{aWW}|$ gives an approximate lower bound to $|\gagg|$.

\subsection{Calabi-Yau hypersurfaces in toric varieties}
In addition to the Tree ensemble, we consider Calabi-Yau hypersurfaces in toric varieties~\cite{1993alg.geom.10003B, Kreuzer:2000xy}. Such a hypersurface combinatorially corresponds to a triangulated reflexive 4d polytope. Systematically orientifolding a large number of Calabi-Yau threefolds is beyond the scope of this work, and so we use the geometric data obtained from the Calabi-Yau itself as a model for the appropriate $\mathcal{N} = 1$ data, namely in constructing $K^{-1}$. 
For the hypersurfaces case, we assume that the gauge group (see the listed options in $\S$~\ref{sec:tree}) is supported on the restriction of a divisor
$Q$ that is a linear combination of toric divisors to the hypersurfaces. There is, {\it a priori}, an infinite number of choices for such a linear combination, but physical considerations will render this set finite and computable, as we will discuss in Sec.~\ref{sec:ens}.

\subsection{Random Matrix EFTs}\label{sec:RMT}

For the sake of comparison to our string results, we will also compute ALP-photon couplings in certain random matrix (RM) effective field theories. 
We emphasize, though, that we do not currently have a reason to believe
that the random matrix ensembles that we study accurately represent actual string data~\cite{Long:2014fba}. Instead, we simply wish to compare and also to demonstrate that they
can also give rise to growing ALP-photon couplings as a function of $N$. This lends some further credence to the idea that $g_{a\gamma\gamma}$ should
increase with $N$.

For our RMT analysis we will consider the simplified Lagrangian of massless ALPs and gauge fields:
\begin{align}\label{eq:rmtlag}
& \mathcal{L}  = -\frac{M_p^2}{2}K_{ij} (\partial^\mu \theta^i)(\partial_\mu \theta^j) -\frac{1}{4}F^{\mu\nu}F_{\mu \nu}  \nonumber \\ 
& - \frac{1}{4}Q_{\, i} \theta^i \tilde{F}^{\mu\nu}F_{\mu \nu} \, ,
\end{align}
where $F_{\mu\nu}$ is the electromagnetic field strength. In a canonically-normalized frame the ALP-photon coupling is simply
\begin{equation}\label{eq:rmtcoup}
\gagg = \sqrt{Q \cdot K^{-1} \cdot Q}\, .
\end{equation}
A natural model for $K$ ($K^{-1}$) is to draw it from an (inverse)-Wishart distribution, as we generally observe that the entries of $K$ ($K^{-1}$) shrink (grow) as a function of  N. Taking $K$ to be a Wishart matrix, we have $K = A^\dagger A$, with the entries of $A$ drawn from a normal distribution $\Omega(0,\sigma)$ centered around zero with standard deviation (SD) $\sigma$. This in turn makes $K^{-1}$ an inverse-Wishart matrix, which in practice is easier to generate directly than to invert $K$. We take $Q$ to be a unit vector, and study two cases: the first being the (unnormalized) entries of $Q$ drawn randomly from the distribution above, and the second taking $Q$ to be a unit vector pointing in a basis direction $Q = \hat{e}_i$.

\section{Distributions of ALP-photon Couplings in the ensembles}\label{sec:ens}
We wish to understand the distribution of $\gagg$ in the generic case, i.e. when $N$ is large. Computing $\gagg$ for the largest $N$ regime of our ensembles in a large number of examples is computationally prohibitive, and so in order to explore the large $N$ regime we will study $\gagg$ for moderately large $N$, and then extrapolate to the largest $N$, where the bulk of the geometries are believed to occur (this has in fact been demonstrated explicitly in a cosmological model on the Tree ensemble, and there is strong evidence for this in the case of hypersurfaces~\cite{Altman:2018zlc}). We will find that in fact the relevant statistical quantities in the distributions of $\gagg$ obey nice scaling properties.

\subsection{The EFT computation}
\subsubsection{Constructing the ensembles}
To construct the ensemble of geometries to study we will randomly draw geometries from the various ensembles. For the Tree ensemble, we construct 1000 geometries for one through ten trees each, corresponding to $h^{1,1}(B)$ ranging from 55 to 235, differing by jumps of $\Delta h^{1,1}(B) = 20$. For a fixed number of trees we randomly draw a tree configuration from the ensemble, and blowup a randomly drawn toric point with that sequence of blowups. The configuration of blowups fixes the topological properties of $B$. 

For the case of hypersurfaces, we randomly draw reflexive 4d polytopes from the Kreuzer-Skarke ensemble~\cite{Kreuzer:2000xy} and compute the pushing triangulation of each polytope to calculate the relevant topological data for the corresponding Calabi-Yau hypersurface. We select 1000 geometries for $h^{1,1}(B) = 10,20,30,40,50,80,120,160$. For $h^{1,1} = 200$ there are only 706 polytopes; we utilized all of them.

Having calculated the relevant topological data, in order to write down the effective field theory for the ALPs and gauge sectors, we need to specify the VEVs of the $\tau^i$. This is done by choosing a point in the K\"ahler cone.
\subsubsection{The K\"ahler cone}
In a complete calculation, the VEVs of the $\tau^i$ are determined by moduli stabilization. We make the assumption that the $\tau^i$ are stabilized in a regime of non-perturbative control. In particular, we require the volumes of all curves to be greater than or equal to unity. This region is known as the stretched K\"ahler cone, and as observed in~\cite{Demirtas:2018akl}, is very narrow at large $h^{1,1}(B)$. As in~\cite{Halverson:2019kna}, where the Tree ensemble was explored in the context of axion reheating, we evaluate the $\tau^i$ at the apex of the stretched K\"ahler cone, defined by minimizing the sum of the toric curves. Since the cone is narrow, and the couplings in Eq.~\ref{eq:gcoups} are invariant under scaling out in the cone via $J \rightarrow \lambda J$, we expect this point to be a good representative of the physics. In addition, as one scales out  $J \rightarrow \lambda J$ the cycle volumes increase, and so more ALPs will become light, which would enhance the couplings of the gauge groups to the ultralight ALPs, and so performing this analysis should provide a lower bound on the size of the couplings in the stretched K\"ahler cone.

\subsubsection{ALP masses and gauge couplings}
 With the EFT data in hand, we proceed to compute $\gagg$. Since we are interested in the effects of the ultralight ALPs we only need to determine which ALPs will have masses much less than any relevant experimental energy, which we take to be $10^{-12}\, \mathrm{eV}$. Recall that for any ALP $\theta^i$, any term in the potential is accompanied by an exponentially suppressed prefactor $\mathrm{exp}(-2\pi \tau^i)$, and therefore the $\theta^i$ who have $\tau^i \gg 1$ will have very small masses. In the full $\mathcal{N} = 1$ SUGRA potential such terms are accompanied by prefactors involving inverse powers of $\mathcal{V}$, the constant term $W_0$ arising from the GVW flux superpotential~\cite{Gukov:1999ya}, and the various two and four-cycle volumes, and such prefactors are generally $\ll \mathcal{O}(1)$ in the stretched K\"ahler cone (see~\cite{Demirtas:2018akl}). In general, linear combinations of divisors can contribute to the superpotential, so to give an upper bound for the ALP mass one should consider a generating set of divisors such that all effective divisors can be expressed as non-negative integer linear combinations of that set. For a toric variety the toric divisors generate the cone of effective divisors, known as the effective cone.\footnote{In the hypersurface case we work under the assumption that the effective cone of the ambient space provides a good approximation to the effective cone of the hypersurface.} We can therefore give an upper bound on the mass scale generated for an ALP $\theta^i$ of the form
\begin{align}\label{eqn:sugra}
& m_i^2 \lesssim \frac{1}{f_{\mathrm{min}}^2}e^{-2\pi \tau^i/C_i}\, ,
\end{align}
where $f_{\mathrm{min}}$ is the smallest ALP decay constant (square root of the smallest eigenvalue of $K$), the $\tau^i$ are the volumes of the generators of the effective cone, and $C_i$ are geometry dependent constants
that range from $1$ to $30$ and are often dual Coxeter numbers. The inverse factor of $f_{\mathrm{min}}^2$ provides the weakest upper bound of the canonical normalization effects. In practice, many $\theta^i$ have their corresponding $\tau^i \gg 1$, and so we can treat such ALPs as massless. Concretely, this means that we compute the upper bounds on the masses using \eqref{eqn:sugra} and the values of $f_{\text{min}}$ and $\tau_i$ in a geometry, and call the ALP $\theta_i$ massless if the associated bound is below a fixed mass threshold. For instance, in the Tree ensemble  the fraction of axions with masses $\leq 10^{-12} \, \mathrm{eV}$ grows from $0.37$ at $N = 55$ to $0.46$ at $N = 195$, at which it becomes approximately constant.

Keeping only these nearly-massless ALPs, i.e. those with mass upper bound below $10^{-12}\, \text{eV}$, we canonically normalize the ALPs and gauge fields. Not every gauge group is a viable candidate for the SM as mentioned in \ref{en:options}. In particular, the gauge coupling in the UV must be large enough to produce the correct low-energy gauge couplings. For a gauge group supported on a cycle given by $Q_i$, we have the relation 
\begin{equation}
\frac{1}{g_{\mathrm{UV}^2 }}\simeq Q_i \tau^i\ ,
\end{equation}
and so any cycle with $Q_i \tau^i$ very
large gives rise to very weak gauge couplings
that are not consistent with well-studied
models and the observed gauge couplings.  For instance,
the SUSY GUT value is $\alpha_{\text{UV}}\simeq .03$, which corresponds to $Q_i\tau^i=2.7$. Leaving some room for 
model building, we demand that $\alpha_{UV}$ is not more than one order of magnitude
smaller than the SUSY GUT value,
and therefore we impose
the cutoff $Q_i \tau^i \leq 25$. In particular, in the hypersurface ensemble we study SM candidates on linear combinations $Q_i$ of toric divisors whose volume satisfies $Q_i \tau^i \leq 25$,  while in the Tree ensemble we study SM candidates on toric divisors with NHCs with volume $\leq 25$.

For each gauge group satisfying this condition, we then compute the coupling of the corresponding ultralight ALP. We present the results in the next section.

\subsubsection{ALP decays to other sectors}

A generic ALP in our setup couples to a large number of gauge sectors (other than the visible sector) via terms of the form $c_{i,\alpha} \phi^i \,\tilde{G}^{\mu\nu}_\alpha G_{\alpha\mu\nu}$, for fixed $\alpha$ (note that this coupling is written for canonically normalized fields). An experimental concern
arises if the ALP decays too quickly
to another sector; for instance, in helioscope experiments the ALPs may in principle decay into a dark sector before it travels from the sun to Earth. The decay rate of an ALP of mass $m_i$ to a pair of dark gluons in one of these sectors is given by 
\begin{equation}
\Gamma_{\phi^i \rightarrow g g} = \text{dim}(G) \frac{c_{i,\alpha}^2 m_i^3}{64 \pi} ~.
\end{equation}
The coupling may be written
\begin{equation}
c_{i,\alpha}\approx (16.6 ~\text{GeV}^{-1}) \times \text{dim}(G)^{-1/2} \left(\frac{1 \,\text{eV}}{m_i}\right)^{3/2} \left(\frac{8 \, \text{min}}{\Gamma^{-1}_{\phi^i \rightarrow g g}}\right)^{1/2}~.
\end{equation}

\noindent Therefore, the coupling has to be $\mathcal{O}(10)$ GeV${}^{-1}$ for an eV ALP produced in the Sun to decay before it reaches the Earth. In all of our ensembles,
the couplings of ALPs to any of the gauge sectors are orders of magnitude less than
$10^{-1}\mathrm{GeV}^{-1}$, and therefore premature decays of ALPs will not be a concern. The ALPs will not decay before reaching the Earth.

\subsection{Results}
\subsubsection{Results for the Tree ensemble}

\begin{figure}[t]
\includegraphics[width=.5\textwidth]{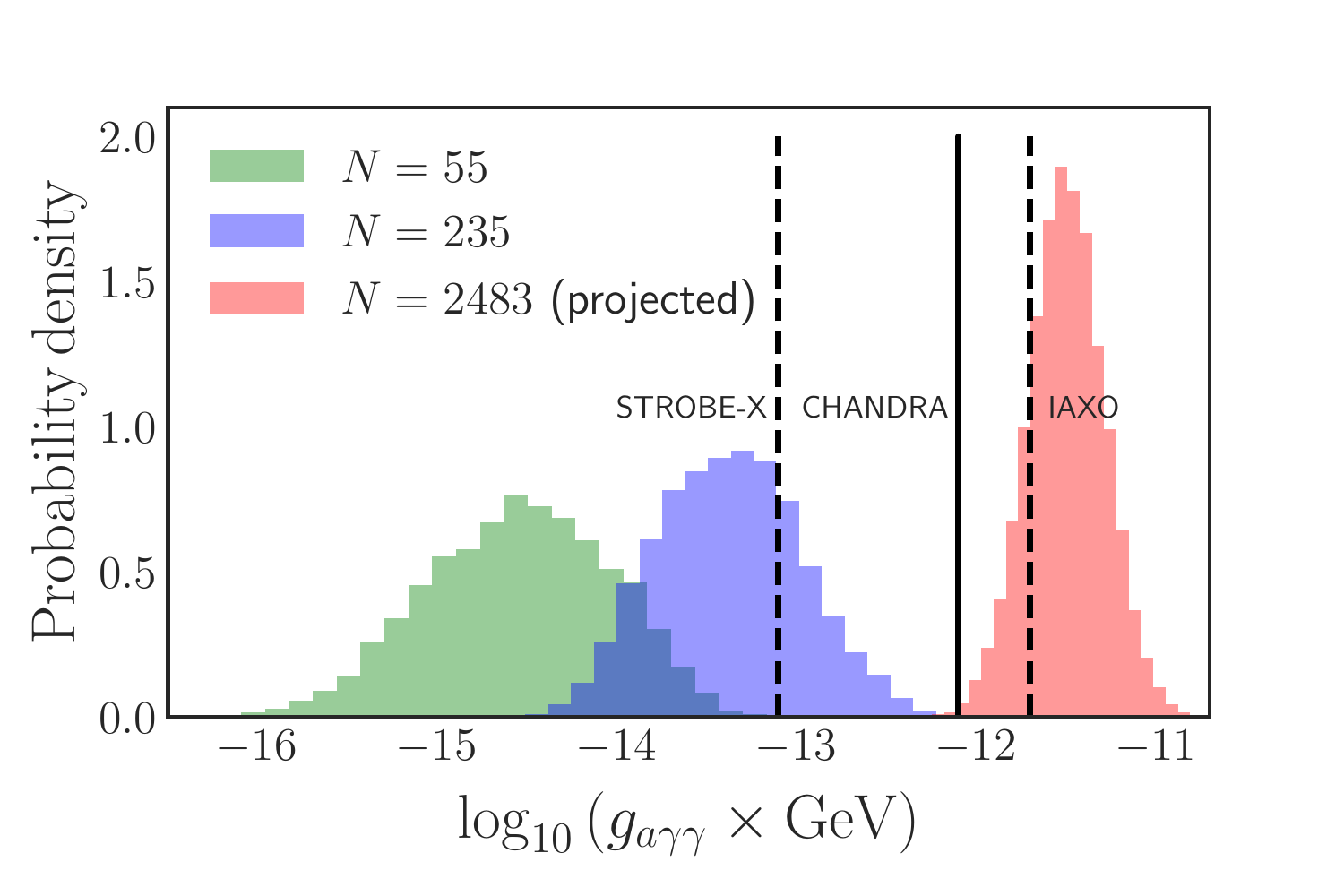}
\caption{
The normalized distributions of $\mathrm{log}_{10}(\gagg  \times \text{GeV})$ for our smallest $N = 55$  and largest $N = 235$, as well as the extrapolated distribution for the preferred value of $N = 2483$ in the Tree ensemble. There is a clear shift of the distribution towards larger values as $N$ grows. Current (solid) and projected (dashed) exclusion lines are presented for various experiments.}
\label{fig:largeN}
\end{figure}

In Fig.~\ref{fig:largeN} we show the normalized distributions of $\mathrm{log}_{10}(\gagg \times \text{GeV})$ for our smallest $N = 55$  and largest $N = 235$. Clearly the distribution shifts to the right as $N$ grows. 

\begin{figure}[t!]
    \centering
    \subfloat[]{{\includegraphics[width=.4\textwidth]{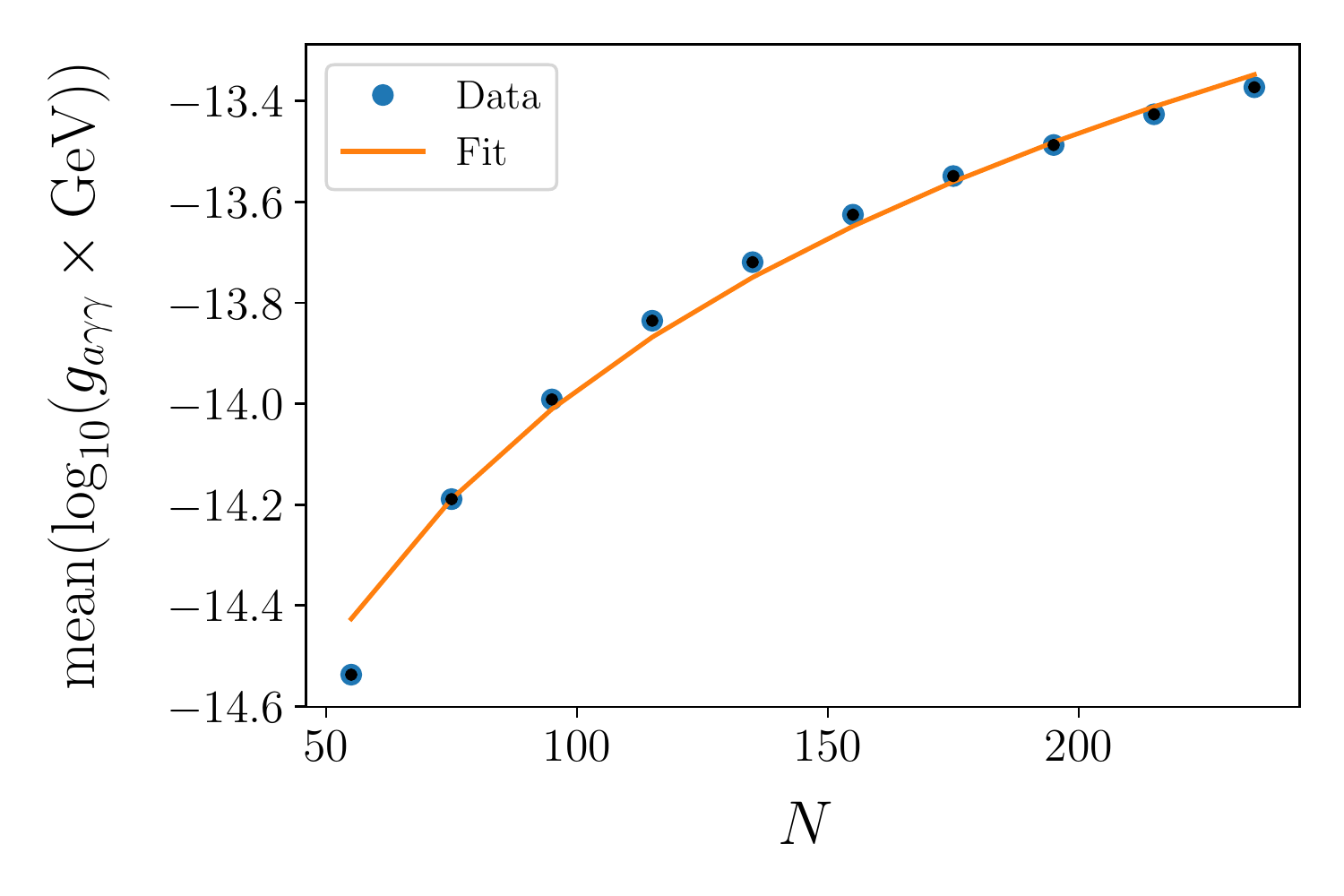}}}
        \qquad
    \subfloat[]{{\includegraphics[width=.4\textwidth]{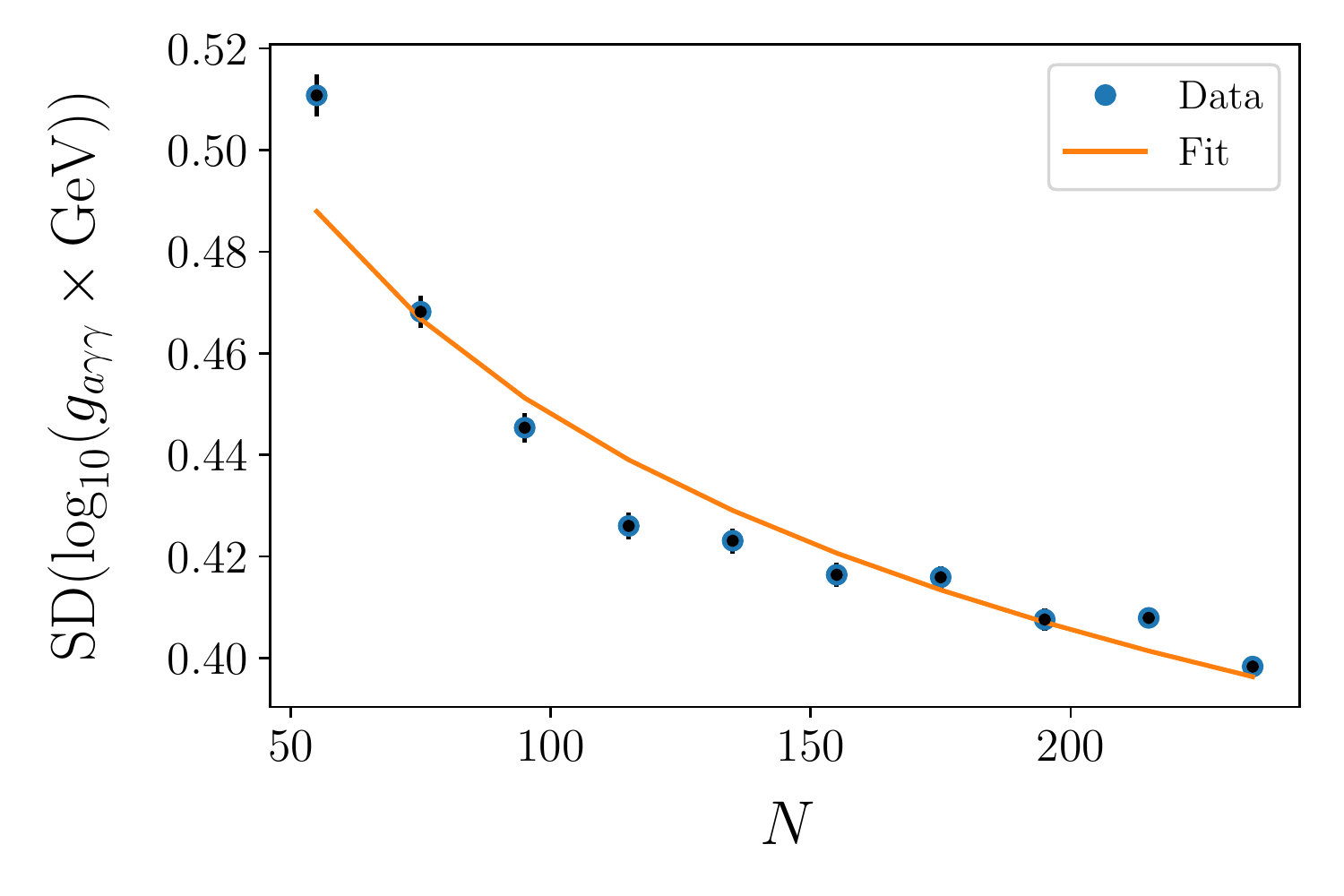}}}
    \caption{
\emph{Top:} The mean of $\mathrm{log}_{10}(\gagg  \times \text{GeV})$ in the Tree ensemble as a function of $N$, with a best-fit curve. \emph{Bottom:} The SD of $\mathrm{log}_{10}(\gagg  \times \text{GeV})$ in the Tree ensemble as a function of $N$, with a best-fit curve.}
    \label{fig:thefits}
\end{figure}

In order to extrapolate to even larger values of $N$, which is often argued to be 
the location of the most vacua in the string landscape, we will determine quantitative properties of the $N$-dependence of the distribution. In particular, we find that both the mean and standard deviation (SD) of $\mathrm{log}_{10}(\gagg  \times \text{GeV})$ have well-behaved $N$-scaling, as shown in Fig.~\ref{fig:thefits}, and are therefore suitable to use for such an extrapolation.

We fit a power-law to the mean of  $\mathrm{log}_{10}(\gagg \times \text{GeV})$ as a function of $N$. Expressed in terms of $\gagg$, we find
\begin{equation}
\text{mean}(\gagg)= 2.73 \times 10^{-18} \times N^{1.77}\,\,\, \text{GeV}^{-1}.
\end{equation}
which gives
the dependence of the coupling on the number
of ALPs. We note here that ``mean'' indicates that we have fit the mean of $\mathrm{log}_{10}(\gagg \times \text{GeV})$, as opposed to $\gagg$ itself.
Given the excellent fit, it is reasonable to extrapolate the mean of the distribution of  $\mathrm{log}_{10}(\gagg \times \text{GeV})$ to the preferred value of $N$ in the Tree ensemble, at $N = 2483$. At this value the predicted mean of $\mathrm{log}_{10}(\gagg \times \text{GeV})$ is $-11.50$, which is much larger than the mean at small-to-moderate $N$. In order to estimate the distribution itself at $N = 2483$ we assume that the distribution is modeled by a Gaussian, with the mean and standard deviation obtained by extrapolating the curves in Fig.~\ref{fig:thefits} to $N = 2483$. The expected distribution is shown in Fig.~\ref{fig:largeN}, along with its smaller-$N$ counterparts, for the sake of comparison.

For the Tree ensemble at large $N$, the distribution sits right on the edge of experimental sensitivity, depending
slightly on the assumptions and experiment. We will present a thorough analysis of our results relative to experimental results and prospects in the discussion, since it will be useful to also compare results across ensembles.

\begin{figure}[t]
\includegraphics[width=.5\textwidth]{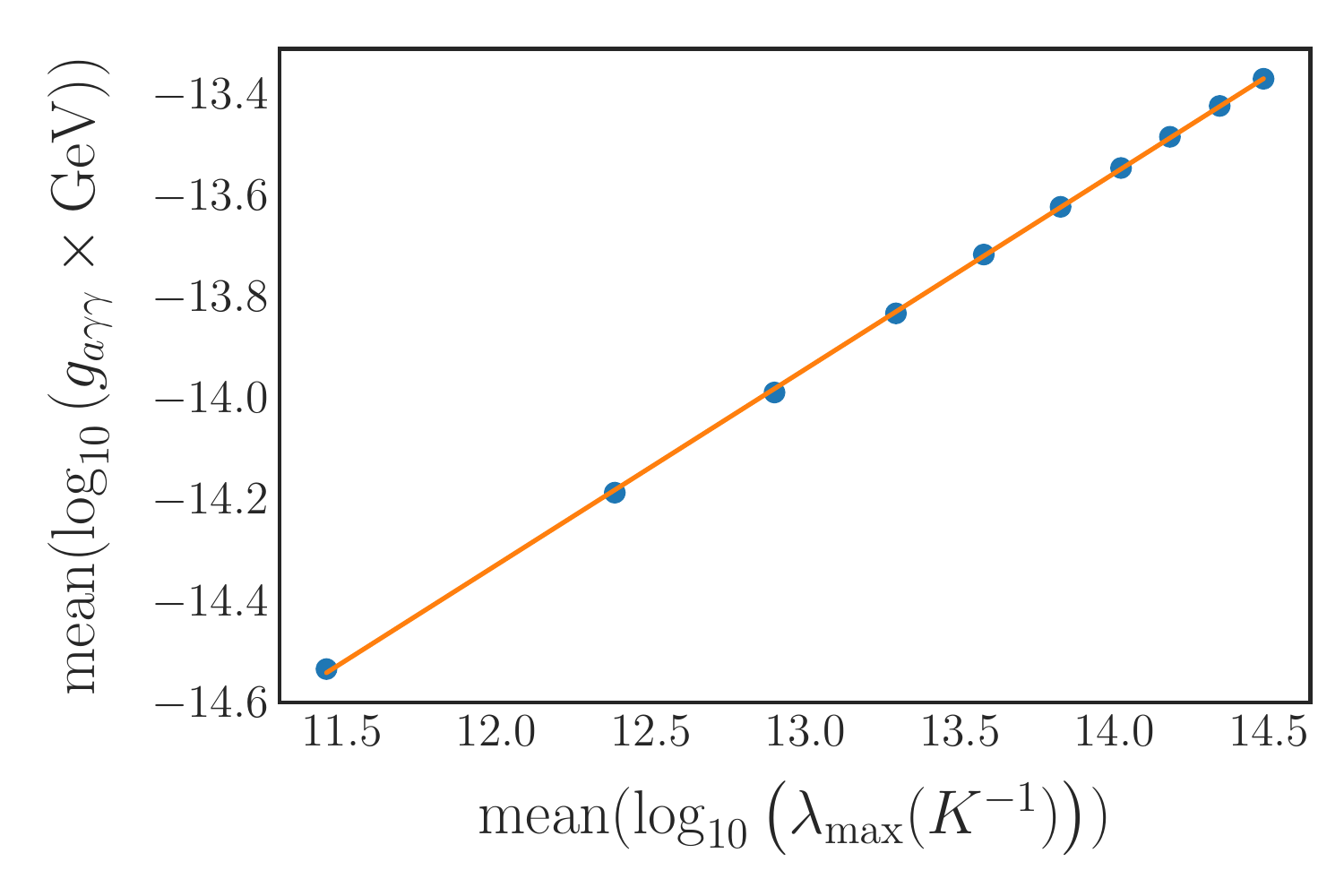}
\caption{
A plot of $\mathrm{mean}(\mathrm{log}_{10}(\gagg  \times \text{GeV})) $ versus $\mathrm{mean}(\mathrm{log}_{10}(\lambda_{\mathrm{max}}(K^{-1})))$. The slope of the line is $\sim 0.38$.}
\label{fig:eval}
\end{figure}

\begin{figure}[t]
\includegraphics[width=.45\textwidth]{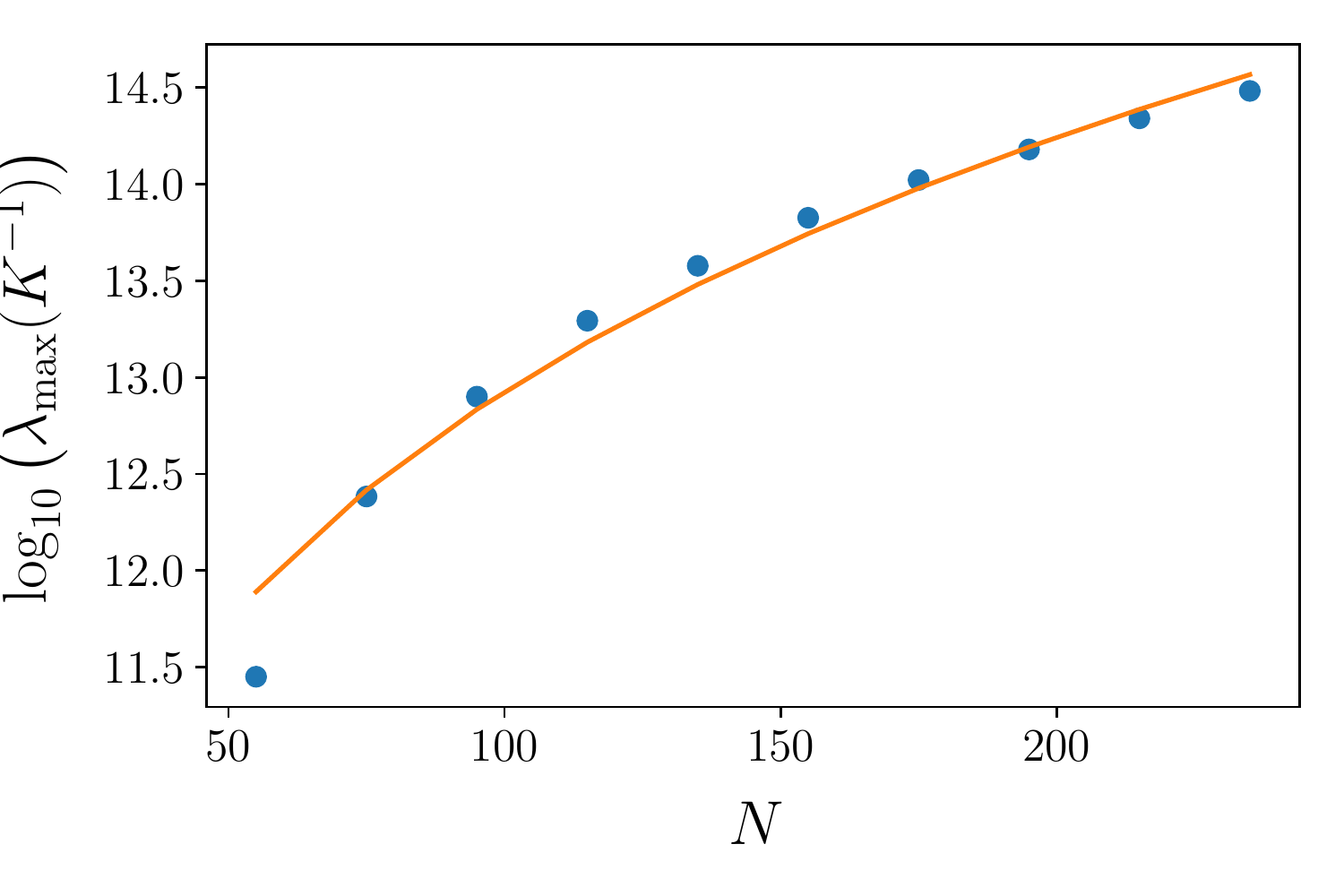}
\caption{
A plot of $\mathrm{mean}(\mathrm{log}_{10}(\lambda_{\mathrm{max}}(K^{-1})))$ versus $N$, with a best-fit line to demonstrate the correlation.}
\label{fig:evalwN}
\end{figure}

\vspace{.5cm}
We would like to understand the origin of this result.
It is clear from our analysis so far, and from Fig.~\ref{fig:largeN}, that as $N$ increases so does the mean of  $\mathrm{log}_{10}(\gagg  \times \text{GeV})$. Recall from Eq.~\ref{eq:gcoups} that $\gagg$ is proportional to the norm of a vector $Q^i$, computed with $K^{-1}$. Since we are imposing $Q\cdot \tau \lesssim 25$ so that the cycle in question can support realistic SM gauge couplings, the denominator of Eq. \eqref{eqn:gaggeqn} is essentially a constant, ranging from $1$ to $25$, compared to the large hierarchies that may arise in the numerator. In Fig.~\ref{fig:eval}, we show a plot of $\mathrm{mean}(\mathrm{log}_{10}(\gagg \times \text{GeV})) $ versus $\mathrm{mean}(\mathrm{log}_{10}(\lambda_{\mathrm{max}}(K^{-1}))) $, where $\lambda_{\mathrm{max}}(K^{-1})$ is the largest eigenvalue of $K^{-1}$. Fig.~\ref{fig:eval} shows a clear correlation, which can be explained as follows: if the largest eigenvalue of $K^{-1}$ grows with $N$, and if $Q^i$ has non-trivial overlap with the corresponding eigenvector that does not shrink too quickly with $N$, then we will find $N$-dependent growth of $\gagg$ with $N$. 
This is indeed the case. In Fig.~\ref{fig:evalwN} we show the $N$-dependence of $\mathrm{mean}(\mathrm{log}_{10}(\lambda_{\mathrm{max}}(K^{-1})))$ with $N$. Clearly $\lambda_{\mathrm{max}}(K^{-1})$ grows rapidly with $N$. In addition, we find that the alignment of the $Q^i$ with the corresponding eigenvector $\hat{v}_\mathrm{max}$ shrinks slowly as a function of $N$: at $N = 55$ we find $\mathrm{mean}(\mathrm{log}_{10}(\hat{v}_\mathrm{max} \cdot \hat{Q})) = -1.72$, while at $N = 235$ we find $\mathrm{mean}(\mathrm{log}_{10}(\hat{v}_\mathrm{max} \cdot \hat{Q})) = -2.59$. Therefore, the growth of $\lambda_{\mathrm{max}}(K^{-1})$ dominates over the decreasing $\hat{v}_\mathrm{max} \cdot \hat{Q}$, explaining the observed correlation.

From a random matrix perspective, constructing a RMT ensemble whose largest eigenvalue reproduces the scaling in Fig.~\ref{fig:evalwN} seems reasonable, as one could choose a Wishart ensemble whose distribution of entries reproduced such a scaling. However, given that $\hat{Q}$ tend to point along the standard basis $\hat{e}$ directions in the natural geometric basis, it is clear that the corresponding eigenvector 
$\hat{v}_\mathrm{max}$ does not obey standard eigenvector delocalization, as we would expect the entries of $\hat{v}_\mathrm{max}$ to be roughly $1/\sqrt{N}$. 

\subsubsection{Results for hypersurfaces}
We perform the same analysis for hypersurfaces, and find quite similar results. In particular, we find that the mean of $\mathrm{log}_{10}(\gagg \times \text{GeV})$ increases as a function of $N$, even more rapidly than in the Tree ensemble, and such an increase can be correlated with the maximal eigenvalue of $K^{-1}$.  

\begin{figure}[t]
\includegraphics[width=.5\textwidth]{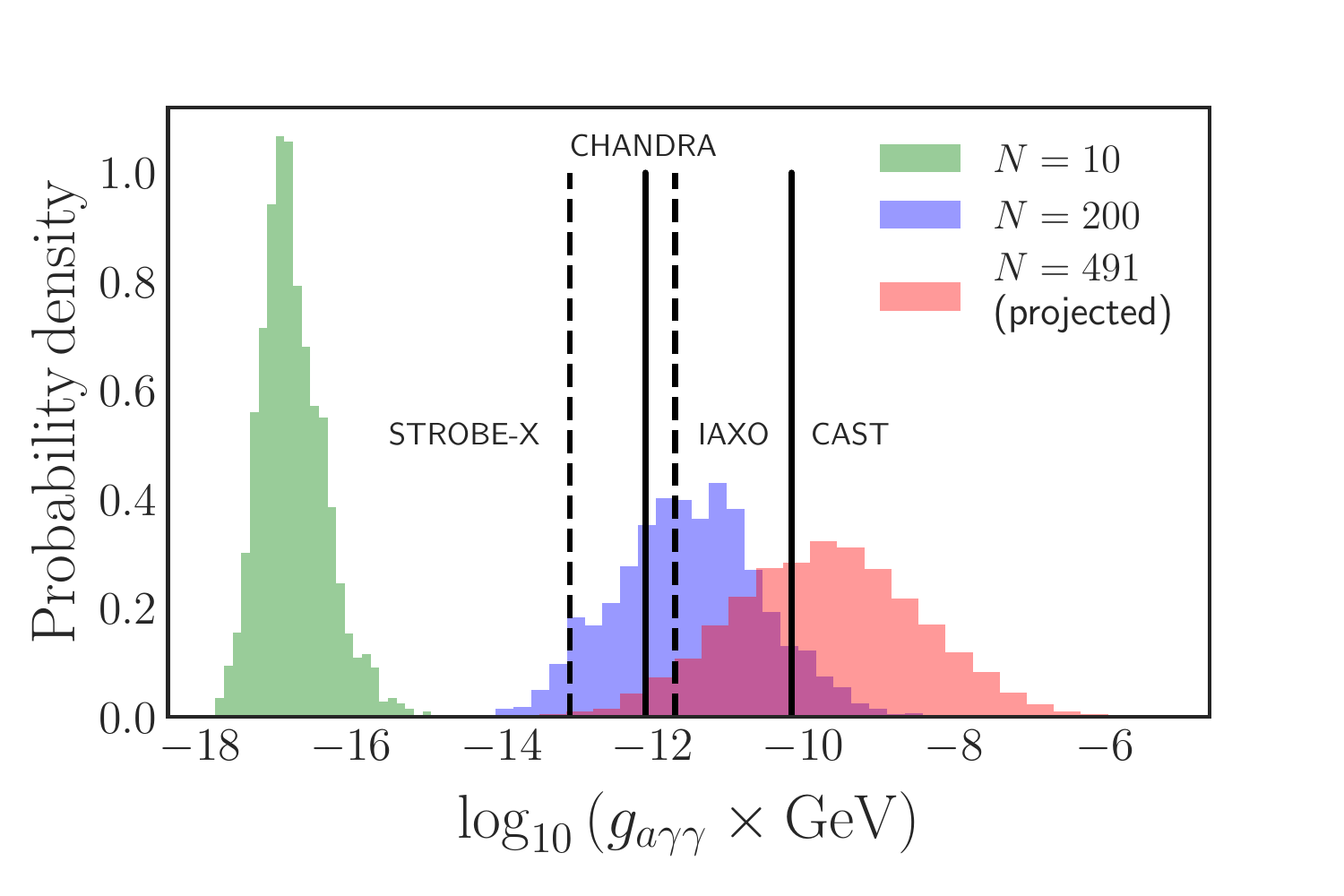}
\caption{
The normalized distributions of $\mathrm{log}_{10}(\gagg  \times \text{GeV})$ for the hypersurfaces. Shown are the calculated distributions for $N = 10$  and  $N = 200$, as well as the extrapolated distribution for the largest hypersurfaces at $N = 491$. There is a clear shift of the distribution towards larger values as $N$ grows. Current (solid) and projected (dashed) exclusion lines are presented for various experiments.}
\label{fig:largeNhyper}
\end{figure}

In Fig.~\ref{fig:largeNhyper} we show the distributions for our smallest and largest $N$ that we analyze in the hypersurface case, which are $N = 10$ and $N = 200$, respectively.  In the same figure we also show the projected distribution for the largest $N$ in the hypersurface ensemble, which is $N = 491$. There is a striking feature in $N$-dependence of the distribution in the hypersurface case as compared to the Tree ensemble: while the standard deviation of the distribution of $\mathrm{log}_{10}(\gagg)$ decreased as a function of $N$ for the Tree ensemble, in the case of hypersurfaces it actually increases, leading to a largest spread for large $N$. This could be due to the rich intersection structure of Calabi-Yau threefolds as compared to toric threefolds, whose intersection numbers are relatively tame.
We fit a power-law to the mean of  $\mathrm{log}_{10}(\gagg \times \text{GeV})$ as a function of $N$. Expressed in terms of $\gagg$, we find
\begin{equation}
\text{mean}(\gagg)= 8.52\times 10^{-22} \times N^{4.22}\,\,\, \text{GeV}^{-1}.
\end{equation}
For the hypersurface at large $N=491$, the mean of the projected distribution
is at $\mathrm{log}_{10}(\gagg \times \text{GeV}) = -9.71$, and so 
a significant portion of the distribution is already in tension with data;
see the discussion below.

\subsubsection{Results for RMT}

As described in Sec.~\ref{sec:RMT} we model $K^{-1}$ as an inverse Wishart matrix (or $K$ as a Wishart matrix) drawn from $\Omega(0,\sigma)$, for some $\sigma$, and $Q$ a unit vector either pointing in a basis direction, or with entries drawn from $\Omega(0,\sigma)$. We find that the different models for $Q$ produce essentially the same behavior, so we will focus on the case that $Q = (1, 0,\dots, 0)$. In the simplified model the coupling of ALP to the photon is given by Eq.~\ref{eq:rmtcoup}. 

\begin{figure}[t]
\includegraphics[width=.5\textwidth]{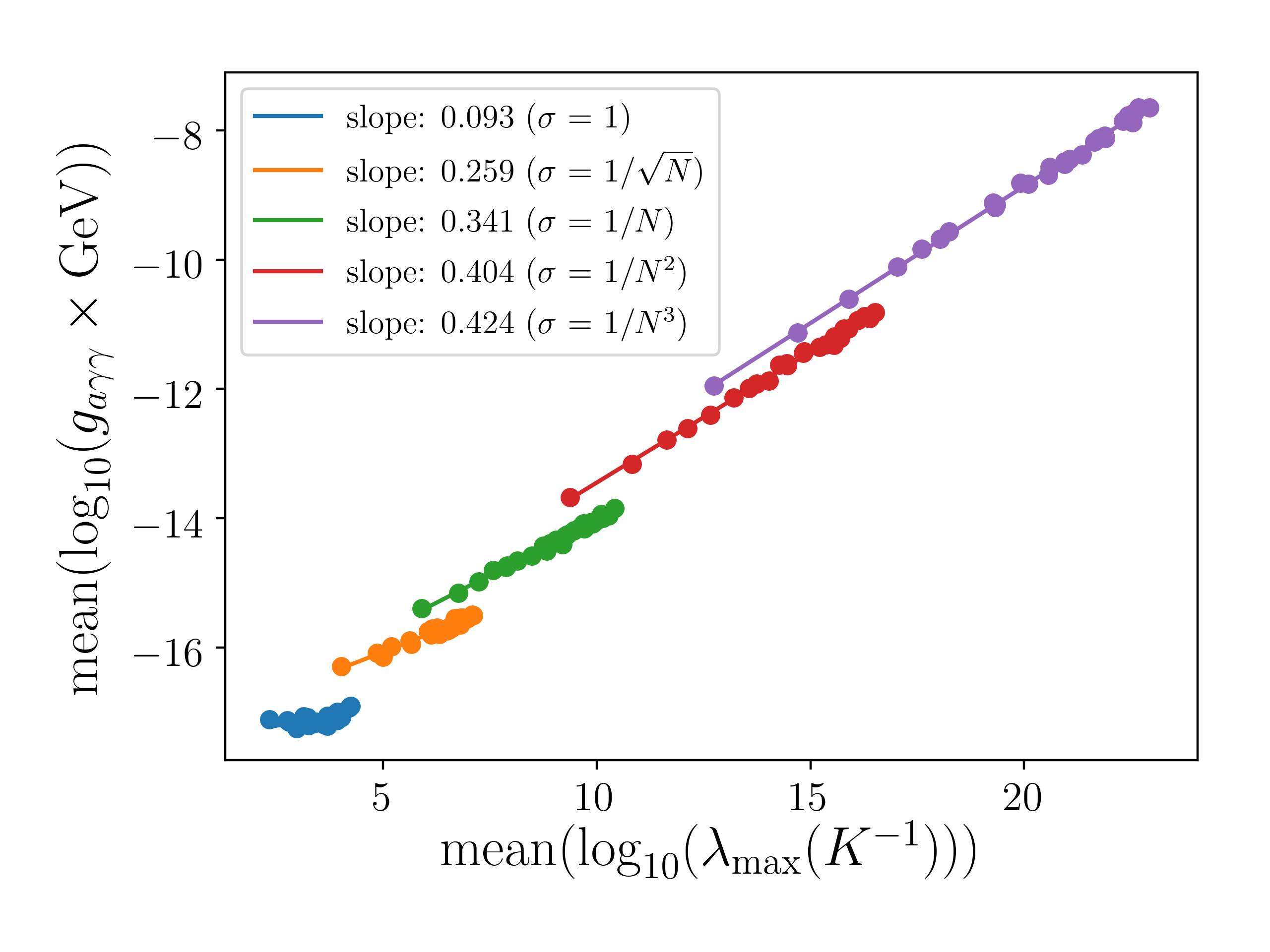}
\caption{
$\mathrm{log}_{10}(\gagg \times \text{GeV})$, computed in the RMT models, versus  $\mathrm{mean}(\mathrm{log}_{10}(\lambda_{\mathrm{max}}(K^{-1})))$, with varying $\sigma$. Note that for all models both $\mathrm{log}_{10}(\gagg  \times \text{GeV})$ and $\mathrm{mean}(\mathrm{log}_{10}(\lambda_{\mathrm{max}}(K^{-1})))$ increase as a function of $N$.}
\label{fig:gooseRMT}
\end{figure}

In Fig.~\ref{fig:gooseRMT} we show $\mathrm{log}_{10}(\gagg  \times \text{GeV})$ computed in the RMT models versus  $\mathrm{mean}(\mathrm{log}_{10}(\lambda_{\mathrm{max}}(K^{-1})))$, to demonstrate the similar behavior to the string ensembles studied above. Implicitly in the plot $\mathrm{mean}(\mathrm{log}_{10}(\lambda_{\mathrm{max}}(K^{-1})))$ is increasing with $N$, and each data point is at a different $N$, ranging from to 0 to 1500. Comparing the growth
of $\text{mean(log}_{10}(\gagg\times \text{GeV}))$ with respect to both $N$ and the maximum eigenvalue of $K^{-1}$, we note that the Inverse Wishart ensembles with $\sigma = 1/N^2$ and $\sigma = 1/N$ are the closest fits to the actual string data, though both differ from it non-trivially.

\begin{figure}[t]
\includegraphics[width=.5\textwidth]{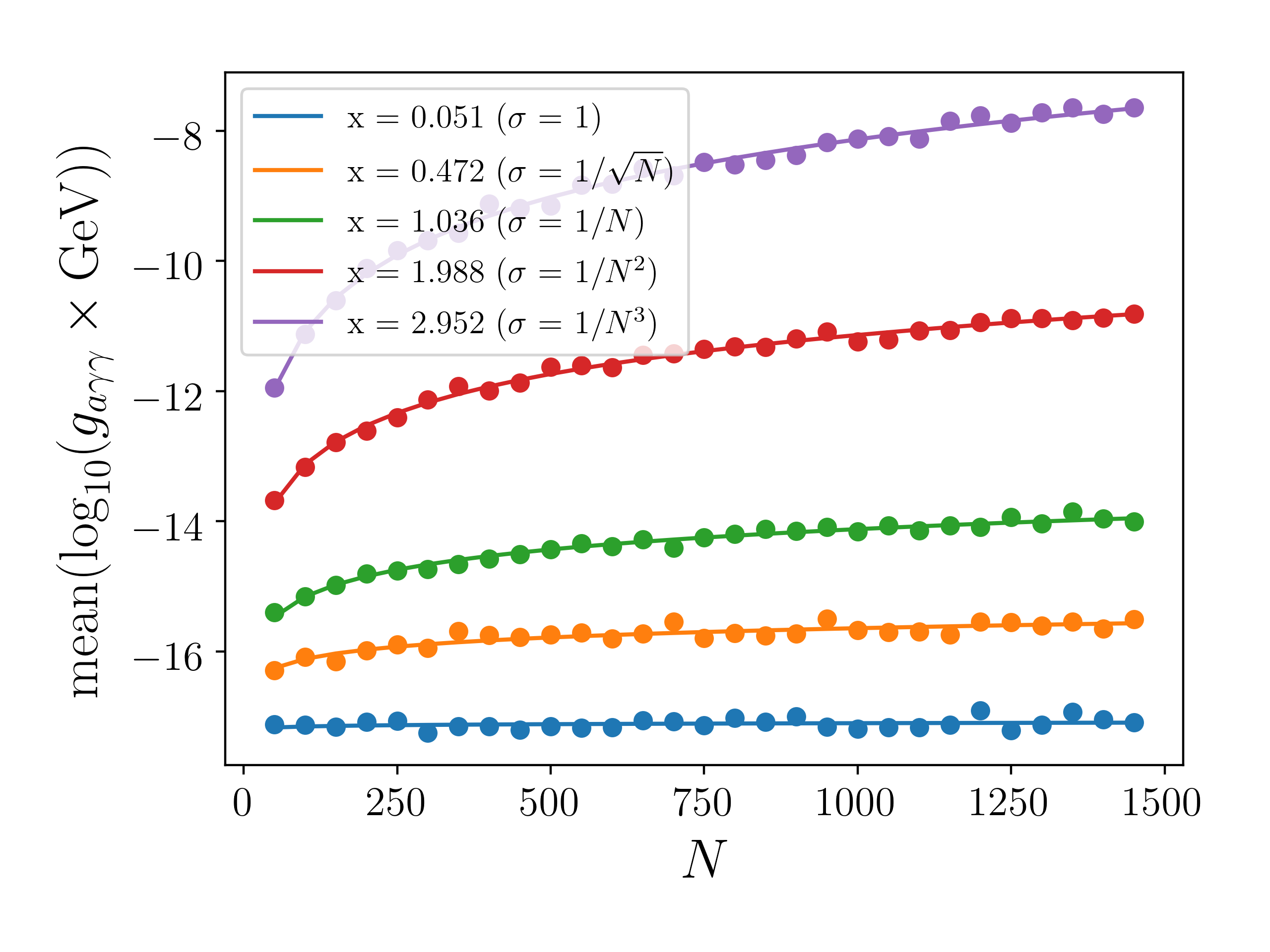}
\caption{The mean of $\mathrm{log}_{10}(\gagg  \times \text{GeV})$ in the Inverse Wishart model for $\sigma = 1, 1/\sqrt{N}, 1/N,1/N^2, 1/N^3$ as a function of $N$, with a best-fit line of the form mean($\gagg$) $\propto N^\mathrm{x}$. The models with $\sigma = 1/N^2$ and $\sigma = 1/N$ are the best for the string data.}.
\label{fig:gN}
\end{figure}

This lends further credence
to the central idea of this work: 
$\gagg$ should grow significantly with
$N$. In particular, if some other string ensemble has a different ALP-photon coupling scaling compared to the ones that we have studied, these EFT considerations suggest that the coupling should nevertheless increase polynomially in $N$, as shown in Fig~\ref{fig:gN}. We therefore believe
that the central result likely extends
to other string ensembles.

\subsubsection{Remarks on other ensembles}
While the Tree ensemble of F-theory bases and the KS ensemble of CY hypersurfaces are the largest-to-date explicit ensembles of string geometries, there are other important examples that we now discuss. 

\begin{figure}[t]
\includegraphics[width=.5\textwidth]{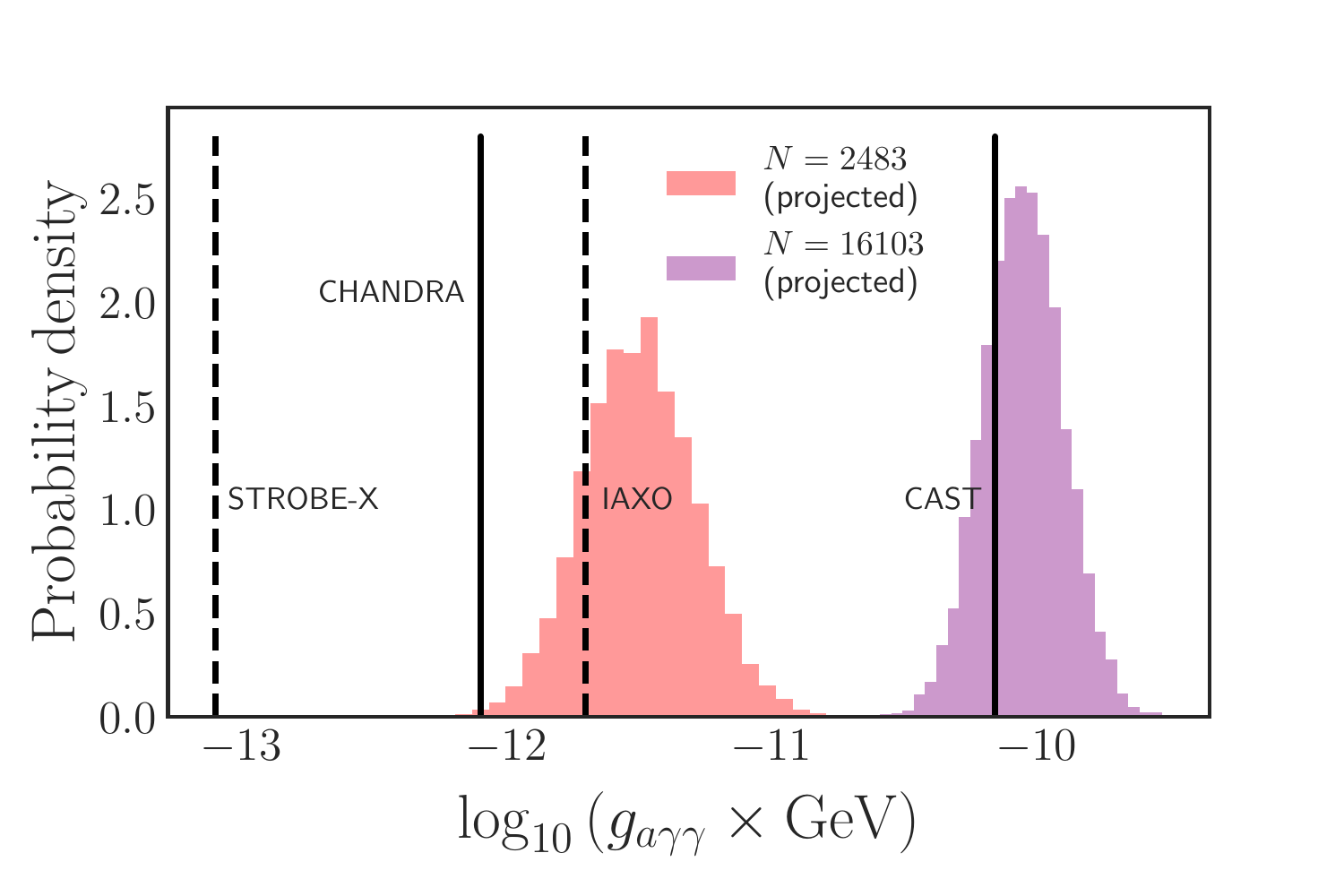}
\caption{
The normalized predicted distribution of $\mathrm{log}_{10}(\gagg  \times \text{GeV})$ for the largest known value of $N = 16103$ in the skeleton ensemble, contrasted with predicted for the largest $N = 2483$ in the Tree ensemble. Current (solid) and projected (dashed) exclusion lines are presented for various experiments.}
\label{fig:skeleton}
\end{figure}

The first is that of the Skeleton ensemble~\cite{Taylor:2017yqr}, which generalizes the Tree ensemble beyond the simple sufficient criteria to remain at finite distance in moduli space. Actually, in this case the construction of the geometries $B$ is a less-restrictive version of the Tree ensemble, and so we expect the results to exhibit similar scaling. However, the largest number $N$ of ALPs found in this ensemble is $16103$, which is a great deal larger than the maximal $N$ found in the Tree ensemble. While this result should be taken with a grain of salt, since we do not perform the
computation for reasons of complexity, it is interesting to check what the distribution of $\gagg$ would be if we were able to extrapolate outside of the Tree ensemble to such an boundary point of the Skeleton ensemble. We extrapolate
using the fit of the related tree
ensemble in Fig.~\ref{fig:skeleton}.

Another important example is the geometry $B_{\mathrm{max}}$, which is the F-theory base thought to house the largest number of flux vacua~\cite{Taylor:2015xtz}. In this geometry the number of ALPs is 98, and the non-Higgsable gauge group is $E_8^9 \times F_4^8 \times \left(G_2 \times SU(2) \right)^{16}$. At the apex of the stretched K\"ahler cone there is only a single cycle with volume $\tau$ that supports a gauge group and satisfies $\tau \leq 25$. This cycle supports an $E_8$ group that could provide a GUT after flux-breaking. For this gauge group we find $\gagg = 3.47 \times 10^{-12}\,\mathrm{GeV}^{-1}$. This coupling is also projected to be probed at future experiments.

\section{Current and Future Experiments}
\label{sec:experiments}

Even though they are not relevant for our analysis, for the sake of thoroughness we will
begin by discussing experiments that
assume that ALPs contribute non-trivially to the dark matter and / or are sensitive only to ALPs in some finite mass region.
Currently, the best limits on the existence of light ($m _a \lesssim 1\,{\rm eV}$) axion-like particles comes from resonant-cavity haloscope experiments. The bounds are restricted to narrow mass-windows centered around $m_a \simeq 2.5 \times 10^{-6}\,{\rm eV}$ for ADMX~\cite{Du:2018uak} and $m_a \simeq 2.5 \times 10^{-5}\,{\rm eV}$ for HAYSTAC~\cite{Zhong:2018rsr}. Both experiments place an impressive limit on the coupling of $g_{a\gamma\gamma} \lesssim 10^{-15}\,{\rm GeV}^{-1}$. However, the effectiveness of these searches is highly contingent on the assumption that the ALP in question comprises the totality of the local dark matter halo, as the signal scales linearly with the local halo density of ALPs. String theory generally provides a wealth of potential dark matter candidates, and even if an ALP were to represent the bulk of the local dark matter density, it would be a remarkable coincidence if this particle was to also couple to photons in an appreciable manner.

Considering the above shortcoming, we are thus led to consider experiments which are both broadband in sensitivity, and independent of assumptions about the ALP contribution to local dark matter. The so-called ``light-shining-through-walls'' (LSW) experiments achieve both, with the DESY-based ALPS~\cite{Ehret:2010mh} and the CERN-based OSQAR~\cite{Ballou:2014myz} achieving a limit of $g_{a\gamma\gamma} \lesssim 6 \times 10^{-8}\,{\rm GeV}^{-1}$ for ALPs with vanishing masses up to a threshold of $m_a \simeq 10^{-4}\, {\rm eV}$. The proposed ALPS-II experiment~\cite{Bahre:2013ywa} hopes to achieve a sensitivity to couplings of order $g_{a\gamma\gamma} \lesssim 2 \times 10^{-11}\,{\rm GeV}^{-1}$ in the very near future. Such a limit will be competitive with the current bound set by the CERN helioscope CAST, which constrains the coupling to $g_{a\gamma\gamma} \lesssim 7 \times 10^{-11}\,{\rm GeV}^{-1}$ for ALPs with vanishing masses up to a threshold of $m_a \simeq 10^{-2}\, {\rm eV}$. This is the current best limit for ALPs in the broad mass range of interest to this paper.

Both of the techniques mentioned above should see improvements in future years. The International Axion Observatory (IAXO), a next-generation helioscope, should reach $g_{a\gamma\gamma} = 2\times 10^{-12}\,{\rm GeV}^{-1}$~\cite{Armengaud:2019uso}, with a prototype taking data within the next few years, while next-generation superconducting radiofrequency cavities hope to reach a similar limit on the same timescale~\cite{Bogorad:2019pbu}.

Finally, we remark on the ability of astrophysical observations to constrain ALP couplings to photons. Oscillations in the spectrum of observed x-ray and gamma-ray photons from extra-galactic sources may be induced by photon-ALP conversion in the presence of large galactic magnetic fields. This technique can be applied for very low ALP masses ($m_a \lesssim 10^{-9}\, {\rm eV}$), with the best limits coming from x-ray astronomy for $m_a \lesssim 10^{-12}\, {\rm eV}$. The most stringent current limits of $g_{a\gamma\gamma} \lesssim 8 \times 10^{-13}\,{\rm GeV}^{-1}$ come from observations of a number of point sources by the Chandra satellite~\cite{Reynolds:2019uqt}. In the next decade, both the European Space Agency (ATHENA)~\cite{Conlon:2017ofb} and NASA (STROBE-X)~\cite{Ray:2019pxr} have proposed missions to improve these limits by an additional order of magnitude.

\section{Discussion} 

\label{sec:discuss}

In this paper we studied distributions of ALP-photon couplings in ensembles of string compactifications and random matrix theories.
Since hundreds or thousands of
axion-like particles typically arise in
string compactifications, 
we studied
the scaling of the effective $\gagg$ distribution
with $N$ and its extrapolation to the
expected value of $N$ in a given ensemble.

Before stating the main results, we review
the setup. 
The details of string geometry, which in our cases are Calabi-Yau manifolds in IIB compactifications or K\" ahler threefold bases of F-theory compactifications, determine the coupling
of a linear combination of ALPs to gauge
sectors arising on a four-manifold 
parameterized by an integral vector $Q$.
Specifically, the string data that enters
the calculation are the choice of manifold from the ensemble, taken to be at the apex of the stretched K\" ahler cone,
the computation of the K\" ahler metric
$K_{ij}$ on ALP kinetic terms that plays a
crucial physical role in canonically
normalized couplings, and choices
of $Q$ that in principle allow for the Standard Model
gauge couplings. ALP masses are generated
by instanton corrections to the 
superpotential, and we utilize a natural
geometric procedure that gives an upper
bound on the ALP mass. The effective
ALP-photon couplings that we study 
include only the contributions from individual
ALPs whose mass upper bound is below
an experimentally relevant threshold $10^{-12}\, \text{eV}$. This conservative choice 
allows our calculations to also apply to 
experiments that are sensitive to ALPs below masses of  $10^{-6}\, \text{eV}$ (e.g., ~\cite{Bogorad:2019pbu}) and 
$10^{-2}\, \text{eV}$ (e.g., helioscopes). Further details
are, of course, provided in the main text.

\vspace{.5cm}

Our main results in each ensemble are
that the mean value of the $\gagg$
distribution ($\overline{\gagg}$) scales polynomially in
$N$, and its extrapolation to large $N$,
where most vacua are argued to occur,
is near the sensitivity of current or 
proposed experiments. We take
each point in turn.

In both string ensembles we studied,
the scaling $\overline{\gagg}$ was significantly stronger
than the $\sqrt{N}$ scaling that one might
naively expect from equation \eqref{eqn:cncoup}. This is due to at least two effects.
First, the number of ALPs contributing
to the effective $\gagg$ depends not only
on $N$, but also the fraction of ALPs with
masses below the chosen threshold; the latter itself increases with $N$ \cite{Demirtas:2018akl}. Second, due to
canonical normalization, $\gagg$ depends
critically on the largest eigenvalue of $K^{-1}$, which
grows significantly with $N$; see Figures \ref{fig:eval} and \ref{fig:evalwN}.
Specifically, the two string ensembles that we 
studied were called
the Tree ensemble and the hypersurface
ensemble, and they exhibited $\overline{\gagg} \propto N^{1.77}$ and $\overline{\gagg} \propto N^{4.22}$, respectively. We note that $\overline{\gagg}$ can also be fit accurately to a weak
exponential, but in the absence of any reason to expect exponential growth,
we opt for the more conservative choice.

In both string ensembles that we studied
we found that $\overline{\gagg}$, extrapolated to the expected large
value of $N$ in each ensemble, is within range
of current and / or proposed experiments.
Specifically, with contributing ALPs having
$m_a\leq 10^{-12} \, \text{eV}$,
\begin{align}
\text{Tree:} \qquad \overline{\gagg} &= 3.2 \times 10^{-12} \, \text{GeV}^{-1} \,\,\,\,\, \text{at} \,\,\,\,\, N =2483 \nonumber  \\
\text{Hyper.:}
\qquad \overline{\gagg} &= 2.0 \times 10^{-10} \, \text{GeV}^{-1} \,\,\,\,\, \text{at} \,\,\,\,\, N = 491.
\end{align}
We also computed $\gagg$ in the F-theory geometry with the most flux vacua, where the requirement $\tau\leq 25$ within the stretched
K\"ahler cone allows the SM to exist only on a single divisor, which carries geometric gauge group $E_8$. In that example, $\gagg = 3.47 \times 10^{-12}\,\mathrm{GeV}^{-1}$.

These results should be compared to the 
strongest current and projected constraints from experiments consistent with the assumptions of our analysis: that the ALPs
do not have to (but could) comprise the dark matter, and that the masses are below a threshold rather than in a fixed window.
For experiments satisfying these assumptions,
the strongest current constraints are
\begin{align}
\text{CAST:} \,\,\,\,\, \gagg &\leq 7.0 \times 10^{-11} \, \text{GeV}^{-1} \,\,\,\,\, \text{for} \,\,\,\,\, m_a \leq 10^{-2} \, \text{eV} \nonumber  \\
\text{Chandra:} \,\,\,\,\, \gagg &\leq 8.0 \times 10^{-13} \, \text{GeV}^{-1} \,\,\,\,\, \text{for} \,\,\,\,\, m_a \leq 10^{-12} \, \text{eV}, 
\end{align}
and the strongest projected sensitivities
in proposed experiments are
\begin{align}
\text{IAXO:} \,\,\,\,\, \gagg &\leq 2 \times 10^{-12} \, \text{GeV}^{-1} \,\,\,\,\, \text{for} \,\,\,\,\, m_a \leq 10^{-2} \, \text{eV} \nonumber  \\
\text{STROBE-X:} \,\,\,\,\, \gagg &\leq 8 \times 10^{-14} \, \text{GeV}^{-1} \,\,\,\,\, \text{for} \,\,\,\,\, m_a \leq 10^{-12} \, \text{eV}.
\end{align}
We note the proximity of the computed $\overline{\gagg}$ in models of ALP-photon
couplings in string compactifications
to the experimental sensitivities.

\vspace{.5cm}
Our study provides the first concrete evidence that ultralight ALPs in 
string compactifications could very well have detectable interactions with photons at large $N$.

Given this strong claim, we would like to clearly emphasize three potential caveats. First,
this is a large $N$ effect, and though there are compelling arguments
that most vacua exist at large $N$, it could be that our vacuum is realized in the small
$N$ regime. Second, the quoted $\overline{\gagg}$ values are the
mean of a projected distribution, and the photon associated with our vacuum could be realized in
the tail. Third, the precise details of the distribution could be modified in more precise
studies that attempt to explicitly realize Standard Model sectors and stabilize moduli, rather
than modeling them, as we have. Each of these caveats is deserving of careful study.

Nevertheless, this result, taken together with the ubiquity of ALPs in string compactifications and the necessity of the photon in the correct theory of quantum gravity, provides excellent motivation for future work.

\vspace{.5cm}

\clearpage
\noindent{\bf Acknowledgements.} We thank Joe Conlon, Caterina Doglioni, Yoni Kahn, Sven Krippendorf,  Liam McAllister, Kerstin Perez, and Ben Safdi for valuable discussions. Portions of this work were completed at the Aspen Center for Physics, which is supported by National Science Foundation grant PHY-1607611. J.H. is supported by NSF grant PHY-1620526 and NSF CAREER grant PHY-1848089. B.N. and G.S. are supported by National Science Foundation grant PHY-1913328.

\bibliography{refs}

\end{document}